\documentstyle[preprint,aps]{revtex}
\def \d {\delta}
\def \ds { \partial \raise.3ex \hbox {\kern -.55 em/}}
\def \ps { p \raise.3ex \hbox {\kern -.55 em/}}
\def \half {{1 \over 2}}
\def\overlay#1#2{\setbox0=\hbox{#1}\setbox1=\hbox to \wd0{\hss
#2\hss}#1\hskip -2\wd0\copy1}
\def\lsim{\mathrel{\rlap{\lower4pt\hbox{\hskip1pt$\sim$}}
    \raise1pt\hbox{$<$}}}         
\def\gsim{\mathrel{\rlap{\lower4pt\hbox{\hskip1pt$\sim$}}
    \raise1pt\hbox{$>$}}}         
\def\beq{\begin{equation}}
\def\eeq{\end{equation}}
\def\bea{\begin{eqnarray}}
\def\eea{\end{eqnarray}}
\def\nn{\nonumber}
\def\eps{\epsilon}

\newcommand\absq{{|\vec q|}}

\newcommand{\bdelta}{\mbox{\boldmath$\delta$}}

\begin{document}
\draft
\preprint{MKPH-T-97-25} 
\title{Optimized $\bdelta$ expansion for relativistic nuclear models}
\author{G. Krein$^{*a}$, R.S. Marques de Carvalho$^{b}, $D.P. Menezes$^b$, 
M. Nielsen$^c$, M.B. Pinto$^b$}
\address{$^a$ Institut f\"ur Kernphysik, Universit\"at Mainz, D-55099 Mainz, 
Germany \\
and \\
Instituto de F\'{\i}sica Te\'orica, Universidade Estadual Paulista, 
Rua Pamplona 145,\\
01405-900 S\~ao Paulo, S.P. Brazil\\
$^b$ Departamento de F\'{\i}sica, Universidade Federal de Santa
Catarina\\
88.040-900 Florian\'opolis, S.C., Brazil\\
$^c$ Instituto de F\'{\i}sica, Universidade de S\~ao Paulo,
Caixa Postal 66318\\
5389-970 S\~ao Paulo, S.P., Brazil}
\maketitle
\begin{abstract}
The optimized $\delta$-expansion is a nonperturbative approach for field
theoretic models which combines the techniques of perturbation theory and the
variational principle. This technique is discussed in the $\lambda \phi^4$
model and then implemented in the Walecka model 
for the equation of state of nuclear matter. The results
obtained with the $\delta$ expansion are compared with those obtained with the
traditional mean field, relativistic Hartree and Hartree-Fock 
approximations. 


\noindent
\vspace{3.0cm}$^*$ Alexander von Humboldt research fellow.

\end{abstract}

\newpage
\hspace*{-\parindent}{\bf 1. Introduction}

The study of possible modifications of hadron properties in the nuclear medium
is one of the central problems of contemporary nuclear physics. In principle, 
these and related phenomena in nuclear physics are governed by the 
fundamental theory of the strong interactions, quantum chromodynamics (QCD). 
However, although QCD has been very successful in explaining a large
class of hadronic processes at high energy and large momentum transfer, typical
nuclear phenomena at lower energies cannot be derived from QCD with the
theoretical tools presently available. The difficulty of using QCD for
phenomena at the nuclear scale is related to the nonperturbative nature
of these. Due to the asymptotic freedom property of QCD, high energy
processes are calculable by perturbative techniques in the quark-gluon
coupling constant. On the other hand, since there are no reliable systematic 
approximation schemes in field theory for performing nonperturbative 
calculations, the construction of models is an important aspect of low energy 
QCD. While there is considerable optimism that eventually one will be able to 
solve QCD numerically on the lattice using supercomputers, the development of 
analytical approximation methods are in urgent need to make contact with the 
wealth of data on nonperturbative phenomena presently available, or that will
be available when the new experimental facilities under construction start 
operating. The $\delta$ expansion~\cite{original} is an example of a method 
recently developed aiming to study nonperturbative phenomena in field theory.

The idea of the $\delta$ expansion is to perturb the original theory by the
introduction of an artificial expansion parameter $\delta$, absent in the
original theory. The parameter $\delta$ is introduced in such a way that it
interpolates between the theory one wants to solve and another theory that
one knows how to solve. The $\delta$ expansion can be formulated in two
different forms, the logarithmic $\delta$ expansion~\cite{original} and the 
linear $\delta$ expansion~\cite{linear}-\cite{su}. In this paper we consider 
the linear form. Specifically, let ${\cal L}$ be the Lagrangian density of 
the theory one wants to solve, and ${\cal L}_0$ the Lagrangian density of the 
soluble theory. Then, the interpolating Lagrangian density  ${\cal L}(\delta)$
is defined as
\beq
{\cal L}(\delta) = (1-\delta){\cal L}_0 + \delta\,{\cal L} = {\cal L}_0 + 
\delta \, ({\cal L}- {\cal L}_0),
\label{sdelta}
\eeq
so that ${\cal L}(0)={\cal L}_0$, ${\cal L}(1)={\cal L}$
and ${\cal L}_0$ is a function of an arbitrary mass parameter $\mu$.
The next step 
involves the evaluation of
desired physical quantities as a perturbation series in powers of 
$\delta$, and then $\delta$ is set equal to $1$ at the end. A 
crucial aspect of the method is the recognition that ${\cal L}_0$ involves 
arbitrary unknown (dimensionful and/or dimensionless) parameters. 

Fixing the arbitrary parameter $\mu$ is the step which brings all 
nonperturbative information contained in the perturbative calculation. Several
ways to fix the arbitrary mass parameter have been proposed in both versions 
of the $\delta$ expansion as well as in the related methods. One physically 
appealing way to fix the unknown parameters, which is the one adopted here, 
is the principle of minimal sensitivity (PMS) introduced in Ref.~\cite{PMS}. 
This variational principle amounts to the requirement that a physical quantity
P($\mu$) should be at least locally independent of these parameters, which 
implies that
\beq
{ \partial {\rm P}(\mu) \over {\partial \mu}}\Big |_{\bar \mu} = 0 ,
\label{pms}\eeq
at $\delta =1$.
The solution to the PMS equation gives $\bar \mu$ as a function of the 
original parameters of the theory including the coupling constant.  
The $\delta$ expansion, together 
with the criterion of the PMS of physical observables, is known as the 
{\em optimized} $\delta$ {\em expansion}. The convergence of the optimized 
$\delta$ expansion has been proved in Ref.~\cite{ian} for a quantum mechanical
problem.

The different forms of the $\delta$ expansion have been successfully applied
to many different problems in quantum mechanics~\cite{QM}, particle 
theory~\cite{part}-~\cite{gromes}, statistical physics~\cite{stat} and lattice
field theory~\cite{du}-~\cite{hugh}. Most applications show that one is always 
able to reproduce traditional non perturbative results  already at lowest 
order in $\delta$. 

Obviously different approximations give different prescriptions so as to 
select a subset of Feynman diagrams among the infinite set which describes a 
physical process and within the $\delta$ expansion this selection is done in 
an essentially perturbative way. Also, as we shall explicitly see, 
the same order in $\delta$ can contain diagrams which
would belong to different orders if we were using other approximations. 
A drawback of traditional nonperturbative analitycal approximations is that 
one has to sum an infinite subset of graphs so as to consider all orders in 
the coupling. This procedure generates problems related to the inclusion of 
higher orders or nonperturbative renormalization or both.   

One advantage of the method presented here is that one deals
with a reduced number of Feynman graphs so that renormalization can be carried
out in a perturbative way before the PMS produces the final finite 
nonperturbative results. Also, because there is no self consistency involved, 
it can be considered more 
economical as far as numerical computations are concerned.
Motivated by these advantages, we have 
recently~\cite{plb} implemented the optimized $\delta$ expansion for the
Walecka model~\cite{wal}. We have investigated vacuum effects by neglecting
exchange diagrams and have shown that the relativistic Hartree approximation
results are exactly reproduced. In a forthcoming work we will present results
which also include exchange diagrams. 

In the present work we do not address the renormalization question by ignoring
vacuum effects. Here only matter effects are considered up to
second order in $\delta$ which includes direct as well as exchange graphs.
The results are compared with
the traditional Hartree and Hartree-Fock approximations. Our aim is just to 
establish the reliability of the method in coping with nuclear matter 
problems. As a byproduct we hope to provide the reader with a powerful 
alternative tool which can be used in investigations aiming to include higher 
order contributions (such as vertex corrections) and vacuum effects.

Before launching into the actual applications a last remark on how to 
implement the optimized $\delta$ expansion is in order. The standard procedure
is to expand the physical quantity of interest ($P$) in orders of $\delta$ 
starting with the interpolated Lagrangian density ${\cal L}(\delta)$. For 
example if $P$ is  the energy density (${\cal E}$) one calculates vacuum to 
vacuum diagrams order by order 
as in perturbation theory using the O($\delta^0$) propagator. 
In this way $\delta$ labels the diagrams contributing 
to ${\cal E}$  and improvements will eventually result from the 
inclusion of higher order terms. Alternatively one can obtain
an exact expression for ${\cal E}$  using the energy-momentum tensor 
($T^{\mu \nu}$) derived from the {\it original} theory ${\cal L}(1)$. 
In this case the final expresion for the energy density is obtained in 
terms of the full propagators, which are then evaluated via the
$\delta$ expansion.

In the present work 
we adopt the latter prescription for the Walecka model. However, the standard 
prescription is being used in a forthcoming work where the energy density is 
derived perturbatively from the generating functional of the interpolated 
theory. We will then be able to check the equivalence between both 
prescriptions.

The paper is organized as follows: in the next section we use the 
$\lambda \phi^4$ model to pedagogically introduce the $\delta$ - expansion
method. In section 3 we apply the $\delta$- expansion to the Walecka
model and in section 4 we present our conclusions.

\vspace{0.5cm}
\hspace*{-\parindent}{\bf 2. The $\lambda \phi^4$ model}

To start with, we consider the scalar $\lambda \phi^4$ theory whose 
Lagrangian density is given by
\beq
{\cal L} = \half (\partial_{\mu} \phi)^2 - \half m^2 \phi^2 - 
{\lambda \over 4!} \phi^4
\eeq
To implement the linear $\delta$ expansion one can consider a general free
scalar Lagrangian density such as
\beq
{\cal L}_0 = \half (\partial_{\mu} \phi)^2 - \half m_0^2 \phi^2 \;\;,
\eeq
where 
\beq
m_0^2 \equiv m^2 + \mu^2
\eeq
$\mu$ being an arbitrary mass parameter. 
Then, according to Eq.~(1) one gets
\beq
{\cal L}(\delta) = \half (\partial_{\mu} \phi)^2 - \half (m^2 + \mu^2) \phi^2 
- \delta ({\lambda \over 4!} \phi^4 - \half \mu^2 \phi^2)
\eeq
The general way the method works becomes clear by looking at the Feynman rules
generated by ${\cal L}(\delta)$. First, the  original $\phi^4$ vertex has its 
original Feynman rule $-i\lambda$ modified to $-i \delta \lambda$. This minor
modification is just a reminder that one is really expanding in orders 
of the artificial parameter $\delta$. Most importantly let us look at the 
modifications implied by the addition of the arbitrary quadratic part. The 
original bare propagator
\beq
i \Delta (p^2) = {i \over {p^2 - m^2 +i\eps}}\,,
\eeq
becomes at zeroth order in $\delta$
\beq
i \Delta (p^2) = {i \over {p^2 - m^2 - \mu^2 +i\eps}}=
{i \over {p^2 - m^2 +i\eps }}\left [ 1 - { (-i\mu^2)i\over {p^2 - m^2 + i\eps}}
\right ]^{-1}\,,
\eeq
or
\bea
i \Delta (p^2) ={i \over {p^2 - m^2 +i\eps}}+
{i \over {p^2 - m^2 +i\eps}}(-i \mu^2){i \over {p^2 - m^2 +i\eps}} \nn \\
+{i \over {p^2 - m^2 +i\eps}}(-i\mu^2){i \over {p^2 - m^2 +i\eps}}(-i\mu^2)
{i \over {p^2 - m^2 +i\eps}} + ...\;,
\label{prop}
\eea
indicating that the term proportional to $ \mu^2 \phi^2$ contained in 
${\cal L}_0$ is entering the theory in a nonperturbative way. On the other 
hand, the piece proportional to ${\delta}\mu^2 \phi^2$ is only being treated 
perturbatively as a quadratic vertex (of weight $i \delta \mu^2$). Since only 
an  infinite order calculation would be able to compensate for the infinite 
number of ($-i\mu^2$) insertions contained in Eq.~(\ref {prop}) one always 
ends up with a $\mu$ dependence in any quantity calculated to finite order in 
$\delta$.

Following the procedure outlined in the Introduction,
the final expression for the quantity $P$ one wants to evaluate
is written in terms of the full propagators which, 
for the $\lambda \phi^4$ theory, is:
\beq
i\Delta ^ *(p^2) = { i \over {p^2 - m_0^2 - \Sigma (p^2) +i\eps}}\,,
\eeq
where $\Sigma (p^2)$ is the self energy. The $\delta$ expansion is then 
implemented via the substitution:
\beq
i\Delta^*(p^2) \rightarrow i\Delta^{\delta}(p^2)={ i \over {p^2 - m_0^2 - 
\Sigma^{\delta} (p^2) +i\eps}}\,,
\eeq
where $\Sigma^{\delta}(p^2)$ is calculated perturbatively in powers of 
$\delta$. This implies $P= P(\mu)$ and the nonperturbative 
results are
obtained by applying the PMS directly to this quantity, as in 
Eq.~(\ref{pms}). 

\vspace{0.5cm}
\hspace*{-\parindent}{\bf 3. Walecka model}

In this section we consider the Walecka model~\cite{wal} for nuclear matter. 
The Lagrangian density of the model is given by
\bea
{\cal L}_{\rm W} &=& \bar \psi\left[\gamma_\mu(i\partial^{\mu}-
g_\omega V^{\mu}) - (M-g_\sigma\phi)\right]\psi +
\frac{1}{2}(\partial_{\mu}\phi\partial^{\mu}\phi -
m_\sigma^2 \phi^2) \nn \\
&-& \frac{1}{4}F_{\mu\nu}F^{\mu\nu}+\frac{1}{2}
m_\omega^2V_{\mu}V^{\mu}\;,
\label{LW}
\eea
where $\psi$ represents the nucleon field operators, $\phi$ and $V_{\mu}$ are
respectively the field operators of the scalar and vector mesons, and
$F_{\mu\nu}=\partial_{\mu}V_{\nu}-\partial_{\nu}V_{\mu}$.

We are interested in the energy density of the system:
\beq
{\cal E} = \frac{1}{V}\int d^3 x \, \left( <\!\Psi|T^{00}|\Psi\!>
- <\!{\rm vac}|T^{00}|{\rm vac}\!>\right),
\label{Edens}
\eeq
where $|\Psi\!>$ is the interacting ground-state of nuclear matter, 
$|{\rm vac}\!>$ is the vacuum state (zero density), and $T^{00}$ is the 
$00$ component of energy-momentum tensor $T^{\mu\nu}$: 
\beq
T^{\mu\nu}_{\rm W}=i\bar\psi\gamma^\mu \partial^\nu\psi+\partial^
\mu\phi
\partial^\nu\phi+\partial^\nu V_{\lambda}F^{\lambda\mu}-
g^{\mu\nu}{\cal L}_{\rm W}\;.
\label{TmunuW}
\eeq
Note that we have not used the nucleon equation of motion. 

Next, we express the energy density in terms of full propagators and full
self-energies~\cite{BLA}:
\bea
{\cal E}_{\rm W} &=& - i \int \frac{d^4k}{(2\pi)^4}\,{\rm Tr}
\left\{ S(k) \left[ \gamma^0k^0 - \left(\gamma^\mu k_\mu - M\right)\right]
\right\} 
- i \int \frac{d^4k}{(2\pi)^4}\,\Delta_\sigma(k)\left[ \frac{1}{2}\left(k^2 - 
m^2_\sigma\right) - (k^{0})^2 \right] \nn \\
&+& i \int \frac{d^4k}{(2\pi)^4}\,\Delta^{\mu}_{\omega \mu}(k)
\left[\frac{1}{2}\left(k^2 - m^2_\omega\right) - (k^{0})^2\right]
+ i \int \frac{d^4k}{(2\pi)^4}\,\left[\Pi_{\sigma}(k)
\Delta_\sigma(k) + \Pi^{\mu \nu}_{\omega}(k) \Delta_{\omega \mu \nu}(k)
\right] \nn\\
&-& <\!{\rm vac}|T^{00}|{\rm vac}\!>,   
\label{Efull}
\eea
where $S(k)$, $\Delta_\sigma(k)$ and $\Delta^{\mu \nu}_{\omega}(k)$ are 
respectively the nucleon, scalar- and vector-meson full propagators, and the 
meson self-energies $\Pi_\sigma (k)$ and $\Pi^{\mu \nu}_\omega (k)$ 
are given by:
\bea
i\Pi_\sigma (k) &=& - i g_\sigma \int \frac{d^4q}{(2\pi)^4}\, {\rm Tr}\left[
S(k+q)\Gamma_\sigma(p+q,q)S(q)\right]\nn\\
&-& (2\pi)^4\delta^4(k)\left\{g_\sigma\int{d^4q\over(2\pi)^4}
{\rm Tr}[S(q)]\right\}^2\,,\label{Pisfull} \\
i\Pi^{\mu \nu}_\omega (k) &=& + i g_\omega \int \frac{d^4q}{(2\pi)^4}\, 
{\rm Tr}\left[\gamma^\mu S(k+q)\Gamma^{\nu}_\omega(p+q,q)S(q)\right]\nn\\
&-& (2\pi)^4\delta^4(k)\left\{g_\omega\int{d^4q\over(2\pi)^4}
{\rm Tr}[\gamma^\mu S(q)]\right\}\left\{g_\omega\int{d^4q'\over(2\pi)^4}
{\rm Tr}[\gamma^\nu S(q')]\right\}.
\label{Pivfull}
\eea
In these equations, the quantities $\Gamma_i,\; i=\sigma, \omega$ are the full 
meson-nucleon vertex functions. These, in turn, are solutions of
Schwinger-Dyson equations that involve higher-order vertex functions (or
scattering T-matrices). The corresponding bare vertices are given by:
\bea
\Gamma_\sigma &=& i g_\sigma\,,\\
\Gamma^\mu_\omega &=& -i g_\omega \gamma^\mu.
\label{barevertices}
\eea

It is worth emphasizing that in Eq.~(\ref{Efull}) direct and exchange
contributions as well as vertex corrections are included, independent
of the order in $\delta$ considered. 

The strategy now is to evaluate the propagators (self-energies) and 
vertex functions according to the  perturbative-variational scheme of the
optimized $\delta$ expansion discussed in the introduction. According to 
Eq.~(\ref{sdelta}), to implement the $\delta$ expansion one needs to 
introduce a ${\cal L}_0$ such that:
\beq
{\cal L}_W(\delta)=(1-\delta){\cal L}_0 + \delta\,{\cal L}_W. 
\label{Ldel}
\eeq
We choose:
\beq
{\cal L}_0=\bar \psi\left(i\gamma_{\mu}\partial^{\mu}-M_0\right)\psi
+\frac{1}{2}(\partial_{\mu}\phi\partial^{\mu}\phi-m_\sigma^2\phi^2)-
\frac{1}{4}F_{\mu\nu}F^{\mu\nu}+\frac{1}{2}m_\omega^2V_{\mu}V^{\mu}\;,
\label{L0Wal}
\eeq
where
\beq
M_0 \equiv M+\mu\;.
\label{M0W}
\eeq
The interpolated Walecka model is then given by:
\beq
{\cal L}_{\rm W}(\d)={\cal L}_0+\delta\left(-g_\omega\bar\psi\gamma_{\mu}
V^{\mu}\psi+g_\sigma \bar\psi\phi\psi+\mu\bar\psi\psi\right)\,.
\label{Walinter}
\eeq

Notice that the $\delta$ expansion technique could have also been applied
to the meson fields explicitly. However, we will eliminate the meson 
propagators in terms of the nucleon propagator, using the exact 
Schwinger-Dyson equations for the meson propagators:
\bea
\Delta_\sigma(k) &=& \Delta_{\sigma 0}(k) + \Delta_{\sigma 0}(k) \Pi_\sigma (k)
\Delta_\sigma(k)\,,\label{scafull} \\
\Delta^{\mu\nu}_\omega(k) &=& \Delta^{\mu\nu}_{\omega 0}(k) + 
\Delta^{\mu\lambda}_{\omega 0}(k) \Pi_{\omega \lambda\sigma} (k)
\Delta^{\sigma \nu}_\omega(k)\,,
\label{vecfull}
\eea
where the meson self-energies are given in 
Eqs.~(\ref{Pisfull})-({\ref{Pivfull}).
In this way, meson effects enter via the nucleon self-energies. This leaves us 
with only one unknown parameter, $\mu$, which will be fixed by the PMS 
condition applied to the energy density. As already discussed, the 
implementation of the method will be  done via the nucleon propagator, which 
depends on the order in $\delta$ considered. 

Notice also that we could have eliminated the meson-nucleon interaction terms
(the terms proportional to the meson self-energies in  
Eq.~(\ref{Efull})) by using the {\em exact} nucleon Schwinger-Dyson 
equation. This would cancel half of the meson kinetic energies~\cite{BLA}.
In the Appendix, we discuss an alternative way to derive the energy 
density~\cite{plb}, appropriate to calculations up to ${\cal O}(\delta^2)$,
in which one eliminates from the beginning the meson field operators in favor
of the nucleon ones.

At zeroth order in $\delta$, the nucleon self-energy, corresponding to the 
interpolated Lagrangian Eq.~(\ref{Walinter}), is obviously zero, i.e.,
$\Sigma^{(0)}=0$. At this order, the single-particle energy is simply given by:
\beq
E(q)=E_0(q)=\left({\vec q}^{2}+M_0^2\right)^{\half}\;,
\eeq
and the nucleon propagator is
\beq
S^0(q)=S^0_F(q)+S^0_D(q),
\label{SFSD0}
\eeq
with
\beq
S^0_F(q)=\left(\gamma^\mu q_\mu+M_0\right)\frac{1}{q^2-M_0^2+i\epsilon},
\label{SF0}
\eeq
\beq
S^0_D(q)=\left(\gamma^\mu q_\mu+M_0\right)\frac{i\pi}{E_0(q)}
\delta\left(q^0-E(q)\right)\theta\left(P_F- | \vec q |\right),
\label{SD0}
\eeq
the Feynman part of the propagator, Eq.~(\ref{SF0}) corresponding to the 
vacuum part and Eq.~(\ref{SD0}) corresponding to the medium part. In what 
follows we do not consider vacuum contributions. 

At this zeroth-order, meson propagators $\Delta_\sigma(k)$ and 
$\Delta^{\mu\nu}_\omega(k)$ of Eqs.~(\ref{scafull})-(\ref{vecfull}) are 
simply:
\bea
\Delta_\sigma(k) &=& \Delta_{\sigma 0}(k) + \Delta_{\sigma 0}(k) \Pi_\sigma (k)
\Delta_{\sigma 0} (k)\,,\label{sca0}\\
\Delta^{\mu\nu}_\omega(k) &=& \Delta^{\mu\nu}_{\omega 0}(k) + 
\Delta^{\mu\lambda}_{\omega 0}(k) \Pi_{\omega \lambda\sigma} (k)
\Delta^{\sigma \nu}_{\omega 0}(k)\,,
\label{vec0}
\eea
with the self-energies $\Pi_\sigma (k)$ and $\Pi^{\mu \nu}_\omega (k)$ given
by:
\bea
\Pi_\sigma (k) &=& - i g^2_\sigma \int \frac{d^4q}{(2\pi)^4}\, {\rm Tr}
\left[S^0(k+q)S^0(q)\right] + i (2\pi)^4\delta^4(k)\left\{g_\sigma
\int{d^4q\over(2\pi)^4}{\rm Tr}[S^0(q)]\right\}^2, \label{Pis0}\\
\Pi^{\mu \nu}_\omega (k) &=& - i g^2_\omega \int \frac{d^4q}{(2\pi)^4}\, 
{\rm Tr}\left[\gamma^\mu S^0(k+q)\gamma^\nu S^0(q)\right]\nn\\
&+& i (2\pi)^4\delta^4(k)\left\{g_\omega\int{d^4q\over(2\pi)^4}
{\rm Tr}[\gamma^\mu S^0(q)]\right\}\left\{g_\omega\int{d^4q'\over(2\pi)^4}
{\rm Tr}[\gamma^\nu S^0(q')]\right\}.
\label{Piv0}
\eea

Hence, we obtain for the zeroth order energy density 
$${\cal E}^{(0)}_{\rm W}={\cal E}^{(0)}_B+{\cal E}^{(0),dir}_{\sigma}+
{\cal E}^{(0),exc}_{\sigma}+
{\cal E}^{(0),dir}_{\omega}+{\cal E}^{(0),exc}_{\omega},$$ 
the following expressions:
\bea
{\cal E}^{(0)}_{B} &=& \gamma\int_0^{P_F}\frac{d^3q}{(2\pi)^3}
\frac{\vec q^2+MM_0}{E_0(q)},
\label{eb0}\\
{\cal E}^{(0),dir}_{\sigma} &=& -\frac{1}{2}\frac{g_\sigma^2}
{m_\sigma^2}\left[\gamma\int_0^{P_F} \frac{d^3q}{(2\pi)^3}\frac{M_0}
{E_0(q)}\right]^2 \label{esd0}\\
{\cal E}^{(0),exc}_{\sigma}
&=& \frac{g_\sigma^2}{2} \gamma\int_0^{P_F}\frac{d^3q}
{(2\pi)^3E_0(q)}\int_0^{P_F}\frac{d^3k}{(2\pi)^3E_0(k)}
\Delta_\sigma([E_0(q)-E_0(k)]^2-(\vec q-\vec k)^2) \nn\\ 
&\times& \left[\left(\frac{1}{2}-1\right)-[E_0(q)-E_0(k)]^2
\Delta_\sigma([E_0(q)-E_0(k)]^2-(\vec q-\vec k)^2)\right]\nn\\
&\times& \left[E_0(q)E_0(k) -\vec q \cdot \vec k +M_0^2 \right],
\label{esx0}\\
{\cal E}^{(0),dir}_{\omega} &=&
\frac{1}{2} \frac{g_\omega^2}{m_\omega^2}
\left[\gamma\int_0^{P_F}\frac{d^3q}{(2\pi)^3}\right]^2\label{evd0}\\
{\cal E}^{(0),exc}_{\omega} &=&
g_\omega^2 \gamma\int_0^{P_F}\frac{d^3q}
{(2\pi)^3E_0(q)}\int_0^{P_F}\frac{d^3k}{(2\pi)^3E_0(k)}
\Delta_\omega([E_0(q)-E_0(k)]^2-(\vec q-\vec k)^2)\nn\\
&\times &
\left[\left(\frac{1}{2}-1\right)-
[E_0(q)-E_0(k)]^2
\Delta_\omega([E_0(q)-E_0(k)]^2-(\vec q-\vec k)^2)\right]\nn\\
&\times & \left.\left[E_0(q)E_0(k)-\vec q\cdot\vec k-2M_0^2\right]\right\}.
\label{evx0}
\eea
In this expression $\Delta_i(k^2), \; i=\sigma,\omega$ is given by:
\beq
\Delta_i(k^2)=\frac{1}{q^2-m_i^2+i\epsilon}.
\label{mesprop}
\eeq
Note that the term proportional to $k^{\mu}k^{\nu}/m^2_\omega$ in the 
vector-meson propagator is dropped due to the conservation of the baryon 
current.

Now we proceed by applying the PMS to
${\cal E}^0_{\rm W}$:
\beq
\frac{d{\cal E}^{(0)}_{\rm W}}{d\mu}=\frac{d{\cal E}^{(0)}_{\rm W}}{dM_0}
\frac{dM_0}{d\mu}=\frac{d{\cal E}^{(0)}_{\rm W}}{dM_0}=0\,.
\eeq

At zeroth order in $\delta$, one can see from Eqs. (\ref{Ldel}) and 
(\ref{L0Wal})
that no interaction between mesons and nucleons are considered. Thus, 
$\Sigma^{(0)}=0$. On the other hand, $M_0$ depends on $\mu$, vide 
Eq.~(\ref{M0W}) and it is precisely this parameter, fixed by the PMS
condition, which introduces all
the non perturbative information related to the interactions. 
Although we are working at zeroth order in $\delta$,
contributions from direct and exchange terms are included 
in equations (\ref{esd0}) - (\ref{evx0}) above.

Let us first consider the contribution from the direct terms only,
which are given by Eqs. (\ref{esd0}) and (\ref{evd0}). 
Application of the PMS to them
yields the following self-consistency condition for $M_0$:
\beq
M_0=M-\frac{g_\sigma^2 }{m_\sigma^2}\gamma\int_0^{P_F}
\frac{d^3q}
{(2\pi)^3}
\frac{M_0}{E_0(q)}\;.
\label{M00}
\eeq
This is exactly the same self-consistency condition for the effective nucleon
mass obtained by means of the Hartree, or mean-field, approximation.

Now, application of the PMS to the full energy density 
leads to a nonlinear equation for $\mu$, or
equivalently for $M_0$, which is more complicated than the one of
Eq.~(\ref{M00}). To avoid this cumbersome expression, we
have chosen to find the minimum of the energy density numerically.
In Figure~1 we compare the nucleon binding energy, $E/A-M$, obtained by 
using only Eq.~(\ref{eb0}) and the direct contributions from 
Eqs.~(\ref{esd0}) and (\ref{evd0})
(solid line) and coupling constants fixed by
fitting the binding energy and density of equilibrium nuclear matter, with
the full binding energy, keeping the same coupling constants (dotted line).
The value of the coupling constants are $g_s^2=91.64$ and $g_v^2=136.2$.
The masses used in all calculations are $M=939$ MeV, $m_v=783$ MeV and 
$m_\sigma=550$ MeV. We find that the full result coincides with those
obtained in a relativistic Hartree-Fock calculation~\cite{wal,hs} which
we also show for comparison (long-dashed line). Note that the dotted and 
long-dashed lines coincide in the figure. Of course, one could renormalize
the model parameters to reproduce 
the bulk saturation properties of nuclear matter. This would give us
the same coupling constants used in the 
the relativistic Hartree-Fock calculation of Ref.~\cite{hs}.
Therefore, the 
PMS condition on the energy density obtained with the zeroth order propagator
of the Walecka model is also 
equivalent to the usual Hartree-Fock solution. This is indeed a very
interesting result since the self-energy expressions are not present
and therefore only the exchange contributions to the energy density
are enough to reproduce, through the minimization of this expression, the
usual Hartree-Fock result.

In Figure~2 we compare the results for the effective nucleon mass in nuclear 
matter as a function of $P_F$ obtained from $\mu$. From this figure, it is 
clear that the results with the exchange terms and renormalized constants 
coincide with the results obtained by using the direct terms only.

Next we check how the previous results change by dressing the
nucleon propagator up to $O(\delta^2)$. 
For this purpose, we start from the calculation of the self-energy.

For infinite nuclear matter, because of the translational, rotational, parity
and time reversal invariances, $\Sigma(q)$ can be generally written in terms of
the unit matrix and the Dirac $\gamma_\mu$ matrices as follows~\cite{hs}:
\bea
\Sigma(q)&=&\Sigma^s(q)-\gamma_\mu \Sigma^\mu(q) \nn\\
&=&\Sigma^s(q^0,\absq)-\gamma^0\Sigma^0(q^0,\absq)+
\vec \gamma\cdot \vec q \, \Sigma^v(q^0, \absq) \;.
\label{genSig}
\eea
The self-energy to second-order in delta is given by:
\bea
\Sigma^{(2)}(p)=&-&\mu \delta+i\frac{g_\sigma^2 \delta^2}{m_\sigma^2}
\int\frac{d^4q}{(2\pi)^4}{\rm Tr}\left[S^{(0)}(q)\right]-
i\frac{g_\omega^2 \delta^2}{m_\omega^2}\int\frac{d^4q}{(2\pi)^4}
\gamma_\mu {\rm Tr}\left[\gamma^{\mu}S^{(0)}(q)\right]
\nn\\
&+&ig_{\sigma}^2 \delta^2 \int\frac{d^4 q}{(2\pi)^4}S^{(0)}(q)
\Delta_{\sigma}[(p-q)^2]-ig_{\omega}^2 \delta^2 \int\frac{d^4 q}{(2\pi)^4}
\gamma_\mu S^{(0)}(q)\Delta_{\omega}[(p-q)^2]\gamma_{\mu}\;,
\label{SigmaW2}
\eea
where the superscript $(2)$ means second order in $\delta$ and 
$\Delta_{\sigma}$ and $\Delta_{\omega}$ are given in Eq.~(\ref{mesprop}).
Again, we have made use of the baryon current conservation. 

It is important to notice that this {\em is not} a self-consistent equation for
$\Sigma^{(2)}$, although its formal similarity with the corresponding 
Hartree-Fock ones~\cite{wal}. The r.h.s. of this equation is expressed 
in terms of functions calculated at the zeroth-order in $\delta$, as is usual 
in a perturbative calculation. 
We evaluate Eq.~(\ref{SigmaW2}) neglecting the Feynman part of the 
nucleon propagator and considering just $S^0(q)$ given by 
Eq.~(\ref{SD0}). Because of this, 
all integrals in Eq.~(\ref{SigmaW2}) are finite and can be easily evaluated.
The first term in Eq.~(\ref{SigmaW2})
comes from the first order contribution in $\delta$ and must be kept at second
order. These expressions are very similar to the ones obtained with the 
Hartree-Fock 
approximation~\cite{hs}. Since there are subtle differences, we write 
them explicitly below. 
\bea
\Sigma^{s(2)}(p) &=& -\delta \mu-\gamma \frac{g_\sigma^2 \delta^2}
{m_\sigma^2} \int_0^{P_F}
\frac{d^3q}{(2\pi)^3}\frac{M_0}{E_0(q)}\nn\\
&+& \frac{1}{4\pi^2p}\int_0^{P_F}dq\;q\;\frac{M_0}{E_0(q)}
\left[\frac{1}{4}g_\sigma^2 \delta^2 \Theta_\sigma(p,q)
-g_\omega^2 \delta^2 \Theta_\omega(p,q)\right], \label{sigs2}\\
\Sigma^{0(2)}(p) &=& -\gamma\frac{g_\omega^2 \delta^2}
{m_\omega^2}\int_0^{P_F}\frac{d^3q}{(2\pi)^3}
- \frac{1}{4\pi^2p}\int_0^{P_F}dq\;q\;\left[\frac{1}{4}
g_\sigma^2 \delta^2 \Theta_\sigma(p,q)+\frac{1}{2} {g_\omega^2} \delta^2
\Theta_\omega(p,q)\right], \label{sig02}\\
\Sigma^{v(2)}(p) &=&-\frac{1}{4\pi^2p^2}\int_0^{P_F}dq\;q\;\frac{q}{E_0(q)}
\left[\frac{1}{2}g_\sigma^2 \delta^2 \Phi_\sigma(p,q)+g_\omega^2 \delta^2 
\Phi_\omega(p,q)\right] \label{sigv2}\;,
\eea
where the functions $\Theta_i(p,q), \Phi_i(p,q), i=\sigma, \omega$, are 
defined by:
\bea
\Theta_i(p,q)&=&\ln \left| \frac{A_i(p,q)+2pq}{A_i(p,q)-2pq}\right|\;,\\
\Phi_i(p,q)&=&\frac{1}{4pq}A_i(p,q)\Theta_i(p,q)-1\;,
\label{ThetaPhi}
\eea
where
\beq
A_i(p,q)=\vec p^2 + \vec q^2 + m_i^2 -\left[E(p)-E_0(q)\right]^2\;.
\label{Apq}
\eeq

One should pay attention to the fact that now the self-energy also 
carries direct and exchange contributions. 

We are in the position to calculate the energy density. We start
by defining the following auxiliary quantities \cite{hs}:
\bea
M^*(q)    &\equiv & M_0+\Sigma^{s(2)}(q)\;,\nn\\
{\vec q}^*   &\equiv & {\vec q}\left[1+\Sigma^{v(2)}(q)\right]\;,\nn\\
E^*(q)    &\equiv & \left[{\vec q}^{*2}+M^{*2}(q)\right]^{\half}\;,\\
\label{auxW*}
q^{* \mu} & = & q^\mu + \Sigma^{\mu(2)} (q)=\left[q^0+\Sigma^{0(2)}(q),
{\vec q}^*\right]\;, \nn
\eea
and writing the nucleon propagator in the
compact form:
\bea
S(q)&=&S_F(q)+S_D(q)\,,\\
\label{SFSD}
S_F(q)&=&\left[\gamma^\mu q^*_\mu+M^*(q)\right]\frac{1}{q^{*\mu}q^*_{\mu}-
M^{*2}(q)+i\epsilon}\,,\\
\label{SF}
S_D(q)&=&\left[\gamma^\mu q^*_\mu+M^*(q)\right]\frac{i\pi}{E^*(q)}
\delta\left(q^0-E(q)\right)\theta\left(P_F-\absq\right)\,,
\label{SD}
\eea
where $E(q)$ is the single-particle energy:
\bea
E(q)&=&\left[E^*(q)-\Sigma^{0(2)}(q)\right].
\label{selfE}
\eea
Note that we have assumed that the nucleon propagator has simple
poles with unit residue. Within the approximation scheme we are working in
this paper, this assumption is satisfied, as can be seen below. 

At this order, for the functions $\Delta_\sigma(k)$, 
$\Delta^{\mu\nu}_\omega(k)$, $\Pi_\sigma (k)$ and  $\Pi^{\mu \nu}_\omega (k)$ 
one has the same expressions as in Eqs.~(\ref{sca0}-\ref{Piv0}), where instead
of $S^0$ one uses the $S$ above. In what follows, vacuum contributions have 
again been neglected. Hence, we obtain for the energy density of nuclear 
matter the following expression:
\beq
{\cal E}_{\rm W}^{(2)}
=\gamma\int_0^{P_F}\frac{d^3q}{(2\pi)^3}\frac{\vec q \cdot \vec q^*+
MM^*(q)}{E^*(q)}+{\cal E}^{(2),dir}_{\sigma}+{\cal E}^{(2),exc}_{\sigma}
+{\cal E}^{(2),dir}_\omega + {\cal E}^{(2),exc}_\omega,
\label{EW}
\eeq
with
\bea
{\cal E}^{(2),dir}_{\sigma} &=& - \frac{1}{2}\frac{g_\sigma^2}
{m_\sigma^2}\left[\gamma\int_0^{P_F}\frac{d^3q}{(2\pi)^3}\frac{M^*(q)}
{E^*(q)}\right]^2 \label{esd2}\\
{\cal E}^{(2),exc}_{\sigma} 
&=& \frac{g_\sigma^2}{2}\gamma \int_0^{P_F}\frac{d^3q}
{(2\pi)^3E^*(q)}\int_0^{P_F}\frac{d^3k}{(2\pi)^3E^*(k)}
\left\{\Delta_\sigma[(q-k)^2] \right. \nn\\
&\times & \left[\left(\frac{1}{2}-1\right)-[E(q)-E(k)]^2
\Delta_\sigma[(q-k)^2]\right]\left[q^{*\mu}k^*_{\mu}+M^*(q)M^*(k)\right],\\
\label{esx2}
{\cal E}^{(2),dir}_\omega &=&
\frac{1}{2}\frac{g_\omega^2}{m_\omega^2}
\left[\gamma\int_0^{P_F}\frac{d^3q}{(2\pi)^3}\right]^2 \label{evd2}\\
{\cal E}^{(2),exc}_\omega
&=& g_\omega^2 \gamma \int_0^{P_F}\frac{d^3q}
{(2\pi)^3E^*(q)}\int_0^{P_F}\frac{d^3k}{(2\pi)^3E^*(k)}
\{ \Delta_\omega[(q-k)^2]\nn\\
&\times & \left[\left(\frac{1}{2}-1\right)-
[E(q)-E(k)]^2\Delta_\omega[(q-k)^2]\right]
\left.\left[q^{*\mu}k^*_{\mu}-2M^*(q)M^*(k)\right]
\right\}.
\label{evx2}
\eea
These expressions are very similar in form to the ones obtained in the
Hartree-Fock approximation. However, one should notice that the self-energies
are not calculated self-consistently as in the Hartree-Fock approximation,
rather they are given by Eqs.~(\ref{sigs2}) - (\ref{sigv2}), which depend
on $M_0$, which by its turn, is determined numerically by minimizing the 
energy density.
Also, differences are
contained in the fermion kinetic energy, the first term in Eq.~(\ref{EW}), 
and in the factors $\left(\frac{1}{2}- 1\right)$ in Eqs. (\ref{esd2}) and
(\ref{evd2}). These differences arise because we are not using the nucleon
Schwinger-Dyson equation. Please refer to the Appendix for an explicit 
derivation in the case one chooses to eliminate the meson field 
operators~\cite{plb} from the beginning. Application of the PMS to the direct 
contributions present in Eqs.~(\ref{esd2}) and (\ref{evd2}),
calculated only with the direct contributions to the self-energies,
yields again the familiar Hartree result, i.e.,
\beq
M^*=M-\frac{g_\sigma^2 }{m_\sigma^2}\gamma\int_0^{P_F}
\frac{d^3q}{(2\pi)^3}
\frac{M^*}{E^*(q)}\;.
\label{M*}
\eeq
From this result it is straightforward to see that
when only the direct terms are considered in the energy density
and self-energies, the mean-field solution is reproduced at any order in 
$\delta$. This result should be compared with the one presented in 
Ref.~\cite{su} where, in the context of the effective potential, it was
found that the $\delta$- expansion and the $1/N$ expansion 
are identical in the large $N$ limit.

For the full energy density, Eq.~ (\ref{EW}) has to be minimized in
terms of $\mu$ and this is done numerically. This is indeed simpler
than the traditional Hartree-Fock procedure, where three coupled
equations (the self-energy expressions) have to be solved self-consistently.

We do not present in Figure~1 the O($\delta^2$) binding energy because it would
be indistinguishable from the HF one. Instead, for comparison purposes, we 
present in Table~1 the results obtained with the self-consistent Hartree-Fock 
approximation and the ones with the $\delta^0$ and $\delta^2$ expansions. We 
note that a simple iterative procedure for solving the Hartree-Fock equations 
do not converge for Fermi momenta larger than $P_F \sim 1.7$~fm$^{-1}$. 
Inspection of the Table reveals the nice convergence towards the Hartree-Fock 
approximations of the results from $\delta^0$ to $\delta^2$. Moreover, one 
sees that in order to reproduce the Hartree-Fock results, it is enough to use 
the simple calculation at zeroth order.

The behavior of $M^*$ 
as a function of the Fermi momentum at this order does not show any noticeable 
difference as compared with the zeroth order results. In Figure~3 we plot 
the energy density $\cal E$ as a function of $\mu$ for $P_F=1.19$ fm$^{-1}$.
The solid line is obtained without the inclusion of the exchange term 
(the PMS solution in this case is given by $\mu/M=-0.275$) and the dashed 
line gives the full second order density energy (the PMS solution is 
$\mu/M=-0.35$). Recall that if one had an exact solution, the energy density
would be independent of $\mu$. In this sense it is gratifying to notice that
$\cal E$ is a very flat function of $\mu$.  This stability in the value of 
the energy density as a function of $\mu$ is very desirable and guarantees 
that even big changes in the value of $\mu$ will not affect physical 
quantities, as the binding energy for instance.

It is important to point out that although we have obtained the
same results for the binding energy within the zeroth and second order
approximations, this is not true at all orders when exchange
terms are included. At fourth order in $\delta$, for example, vertex
corrections will appear and the resulting energy density 
will certainly be different. In this work we have opted for neglecting
vertex corrections in order to be able to compare our results
with Hartree and Hartree-Fock results, where they
are not included either. If vertex corrections are to be included,
the full meson-nucleon vertex functions $\Gamma_i$, $i=\sigma,\omega$
appearing in Eqs. (\ref{Pisfull}) and (\ref{Pivfull})
and the full meson propagators will also have to
be expanded in orders of $\delta$.

\vspace{0.5cm}
\hspace*{-\parindent}{\bf 4. Conclusions }

In the third section of this paper we have utilized the optimized 
$\delta$ expansion to study medium effects in the Walecka model. 
We have obtained results
quantitatively similar to the ones of the usual Hartree-Fock 
approximation, 
although the analytical expressions are not evidently equivalent. 
If one neglects the exchange term in the energy density 
and self-energies then clearly 
the mean-field solution is reproduced at any order. This outcome 
reflects the fact that this perturbative method generates 
nonperturbative results due,
of course,of the variational nature of the PMS. 

Figure 1 and Table~1 show that the very simple calculation at zeroth order
in $\delta^0$  already provides a very good approximation to the Hartree-Fock 
results, at least for densities not much higher than the normal nuclear matter
density. Analytically, this calculation is indeed simpler than the usual 
Hartree-Fock approximation, in view of the perturbative nature of the method.
Numerically, this calculation is straighforward because no self-consistency 
has to be achieved; one needs only to perform a minimization of the energy 
with respect to the parameter $\mu$. It is also worth mentioning that, in the 
Walecka model, the energy  density is a very flat function of $\mu$ and this 
guarantees that the PMS solution is indeed very stable. 

On the basis of our results, we believe that the optimized $\delta$ 
expansion is a very robust nonperturbative approximation scheme. Compared with
the Hartree-Fock approximation, the $\delta$ expansion is very economical 
because of its perturbative nature. Once the reliability of the scheme
has been established, one is ready to proceed to other interesting 
applications. These include vertex and, obviously, vacuum effects that 
include exchange corrections~\cite{plb}.
In view of our results, we can proceed by including vertex corrections 
in the energy density and still maintaining the nucleon propagator  
at zeroth order in $\delta$. The study of the vacuum in the 
Walecka model is an important issue since one needs to know the limits of 
applicability of such model to high densities and/or temperatures before 
quark and gluon degrees of freedom have to be invoked. Particularly interesting
is the renormalization of exchange diagrams which should become simplified
in the present approach as compared with the Hartree-Fock 
scheme~\cite{SerBiel}, since at each order in $\delta$, only a finite
number of diagrams has to be taken into account.

\section{Acknowledgments}
This work was partially supported by CNPq and FAPESP (contract \# 93/2463-2). 
GK is supported in part by the Alexander von Humboldt Foundation and FAPESP. 
RSMC is supported by CNPq.
We would like to thank Dr. P. Garcia and Dr. F.F. de Souza Cruz for carefully
reading the manuscript and for their useful suggestions. 
 
%

%

\begin{table}
\caption{Results for $E/A-M$~(MeV) as a function of $P_F$~(fm$^{-1}$)
calculated with the Hartree-Fock ($E_{HF}$) approximation and
with the $\delta^0$ ($E_{\delta^0}$) and $\delta^2$ $(E_{\delta^2}$)
expansions.}
\begin{tabular}{cddd}
$P_F$&$E_{HF}$ &$E_{\delta^0}$ &$E_{\delta^2}$ \\
\tableline
0.05                        & 0.0291095 & 0.0291095 & 0.0291095    \\
0.25                        & 0.5317530 & 0.5317529 & 0.5317530    \\
0.50                        & 1.1637572 & 1.1637562 & 1.1637569    \\
0.75                        & 0.5920830 & 0.5920533 & 0.5920831    \\
1.00                        &-1.8171553 &-1.8175904 &-1.8171548    \\ 
1.25                        &-4.0899150 &-4.0937442 &-4.0898972    \\
1.50                        & 5.7406683 & 5.7109528 & 5.7409897    \\
1.65                        &31.0297464 &30.9322795 &31.0312656    \\
\end{tabular}
\end{table}

\begin{figure}
\caption{$P_F$ dependence of the binding energy of the Walecka model at
zeroth order in $\delta$. The solid line represents the 
direct contributions only (Eqs.(34),(35),(37)). 
The dotted and long-dashed lines give the full binding energy
and the Hartree-Fock solution respectively, both determined with the same
coupling constants used in the solid line solution (note that the lines are
coincident).} 
\label{fig1}
\end{figure}
\begin{figure}
\caption{ Zeroth order nucleon effective mass $M_0$ as a function of $P_F$.
The solid curve is the result obtained without the exchange term
and the dashed curve is the result using the full energy density.}
\label{fig2}
\end{figure}
\begin{figure}
\caption{$ \mu$ dependence of the energy density for the Walecka model at 
second order in $\delta$, calculated at $P_F=1.19$~~fm$^{-1}$. The solid line
gives the solution when the exchange term is not included. The dashed line
gives the full solution.}
\label{fig3}
\end{figure}

\newpage
\begin{center}
{\bf Appendix}
\end{center}

For completeness, in this Appendix we employ the method used in
Ref.~\cite{plb} for obtaining the general expression for the energy density 
in terms of the nucleon propagator, valid up to order ${\cal O}(\delta^2)$.

The energy-momentum tensor is defined by Eq.~(\ref{LW})
with ${\cal L}_{W}$ defined in Eq.~(\ref{TmunuW}):
\beq
T^{\mu\nu}_{W}=T^{\mu\nu}_{B}+T^{\mu\nu}_{\sigma}+T^{\mu\nu}_{\omega}
\eeq
with
\bea
T^{\mu\nu}_{B} &=& i\bar \psi\gamma^\mu\partial^\nu\psi-g^{\mu\nu}\bar\psi
(i\gamma_\alpha\partial^\alpha-M)\psi\; ,\\
T^{\mu\nu}_{\sigma} &=& \partial^{\mu}\phi\partial^{\nu}\phi-g^{\mu\nu}
\left(g_\sigma
\bar\psi\phi\psi-{1\over2}m_\sigma^2\phi^2+{1\over2}\partial^{\alpha}
\phi\partial_{\alpha}\phi\right)\; \label{tmunus},\\
T^{\mu\nu}_{\omega} &=& \partial^\nu V_{\lambda}F^{\lambda\mu}-g^{\mu\nu}
\left(-g_\omega\gamma_\alpha V^\alpha +{1\over2}m_\omega^2V_\alpha V^\alpha 
- {1\over4}F_{\alpha\beta}F^{\alpha\beta}\right)\; .
\eea

Let us concentrate on $T^{\mu\nu}_{\sigma}$. The 
Euler-Lagrange equation for the scalar-meson field equation is:
\bea
\left(\partial_\mu\partial^\mu+m_\sigma^2\right)\phi = g_\sigma
\bar\psi\psi\;.
\label{eqsmes}
\eea
This equation can formally be integrated as:
\bea
\phi(x)&=&\phi^{0}(x)-g_\sigma\int d^4y\Delta_\sigma(x-y)\bar\psi(y)\psi(y)\;,
\label{phisol}
\eea
where $\phi^0$ is the solution of the homegeneous equation
and $\Delta_\sigma(x)$ is given by:
\beq
\Delta_\sigma(x)=\int \frac{d^4q}{(2\pi)^4}\frac{1}{q^2-m_\sigma^2+
i\epsilon}e^{-iqx} 
=\int \frac{d^4q}{(2\pi)^4}\, \Delta_\sigma (q^2)\, e^{-iqx}.
\label{propcoord}
\eeq
From Eq.~(\ref{phisol}) (note that the first term $\phi^{0}(x)$ 
does not contribute) we have:
\beq
g_\sigma<{\overline\psi}\phi\psi>=-g_\sigma^2\int d^4y\Delta_\sigma(x-y)
<\bar\psi_\alpha(x)\bar\psi_\beta(y)\psi_\beta(y)\psi_\alpha(x)>\; .
\eeq
With the help of Wick's contraction technique we get
\bea
g_\sigma<{\overline\psi}\phi\psi>&=&g_\sigma^2\int d^4y\int{d^4p\over(2\pi)^4}
e^{-ip(x-y)}\Delta_\sigma(p^2)\nn \\
&\times &\left[\int{d^4q\over(2\pi)^4}{\rm Tr}[S(q)]
\int{d^4k\over(2\pi)^4}{\rm Tr}[S(k)] \right . \nn \\
&-& \left .\int{d^4q\over(2\pi)^4}
e^{-iq(x-y)}\int{d^4k\over(2\pi)^4}e^{-ik(y-x)}{\rm Tr}[S(q)S(k)]\right]\; ,
\eea
that can finally be written as
\beq
g_\sigma<{\overline\psi}\phi\psi>=-{g_\sigma^2\over m_\sigma^2}
\left[\int\frac{d^4q}{(2\pi)^4}{\rm Tr}[S(q)]\right]^2-g_\sigma^2
\int\frac{d^4q}{(2\pi)^4} \frac{d^4k}{(2\pi)^4}{\rm Tr}\left[S(q+k)S(k)\right]
\Delta_\sigma(q^2)\; .
\eeq

The third term in Eq.~(\ref{tmunus}) above can be written as
\beq
-{1\over2}m_\sigma^2<\phi^2>=-{1\over2}m_\sigma^2g_\sigma^2\int d^4y d^4z
\Delta_\sigma(x-y)\Delta_\sigma(x-z)<\bar\psi_\alpha(y)\psi_\alpha(y)
\bar\psi_\beta(z)\psi_\beta(z)>.
\eeq
Using again Wick's technique
\bea
-{1\over2}m_\sigma^2<\phi^2>&=&{1\over2}m_\sigma^2g_\sigma^2\int d^4y d^4z
\frac{d^4p}{(2\pi)^4}\frac{d^4q}{(2\pi)^4}e^{-ip(x-y)}\Delta_\sigma(p^2)
e^{-iq(x-z)}\Delta_\sigma(q^2)\nn \\
&\times & \left[\int{d^4k\over(2\pi)^4}{\rm Tr}[S(k)]
\int{d^4k'\over(2\pi)^4}{\rm Tr}[S(k')] \right . \nn \\
&-& \left .\int{d^4k\over(2\pi)^4}
e^{-ik(z-y)}\int{d^4k'\over(2\pi)^4}e^{-ik'(y-z)}{\rm Tr}[S(k)
S(k')]\right]\; ,
\eea
and finally
\bea
-{1\over2}m_\sigma^2<\phi^2>&=&{1\over2}{g_\sigma^2\over m_\sigma^2}
\left[\int\frac{d^4q}{(2\pi)^4}{\rm Tr}S(q)\right]^2 \nn \\
&-&{1\over2}m_\sigma^2
g_\sigma^2\int\frac{d^4q}{(2\pi)^4} \frac{d^4k}{(2\pi)^4}\Delta_\sigma(q^2)
{\rm Tr}\left[S(q+k)S(k)\right]\Delta_\sigma(q^2)\; .
\eea

Following the same procedure for $\partial^0\phi\partial^0\phi$ and 
$\partial_\mu\phi\partial^\mu\phi$, one obtains after adding all terms:
\bea
T^{00}_s &=& + {1\over2}{g_\sigma^2\over m_\sigma^2}
\left[\int\frac{d^4q}{(2\pi)^4}{\rm Tr}S(q)\right]^2 
+ g_\sigma^2\int\frac{d^4q}{(2\pi)^4} \frac{d^4k}{(2\pi)^4}
{\rm Tr}\left[S(q+k)S(k)\right]\Delta_\sigma(q^2)\nn\\
&\times &\left[(q^0)^2\Delta_\sigma(q^2)
+\frac{1}{2}(q^2+m^2_\sigma)\Delta_\sigma(q^2)\right]\;\;\;\;\;{}.
\label{ultima}
\eea

The same can be repeated for the vector-meson field.  This shows
the equivalence up to ${\cal O}(\delta^2)$ of the perturbative solution of the
Schwinger-Dyson equations and the method of 
elimination of the meson field equations from the beginning~\cite{plb}.

\end {document}
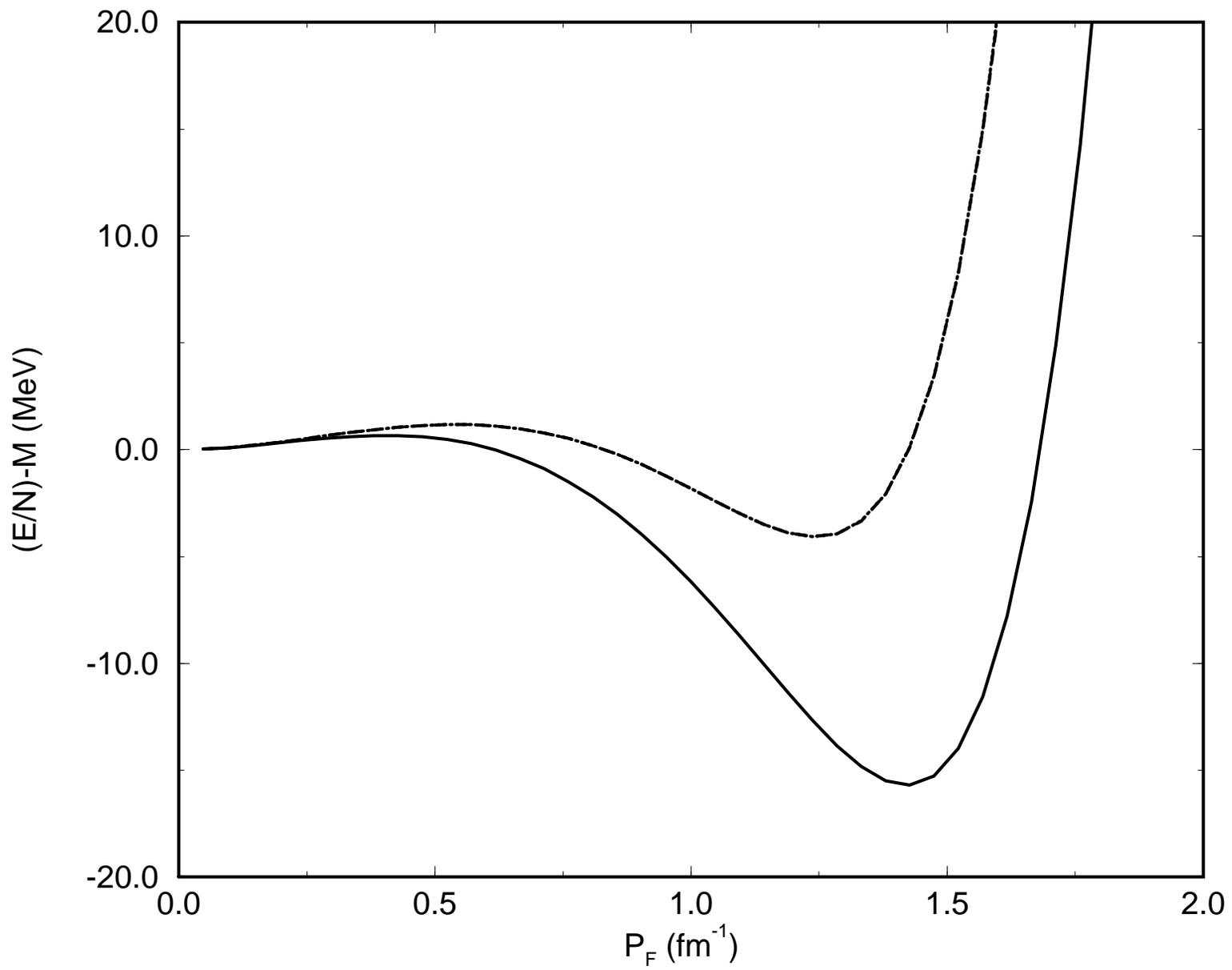
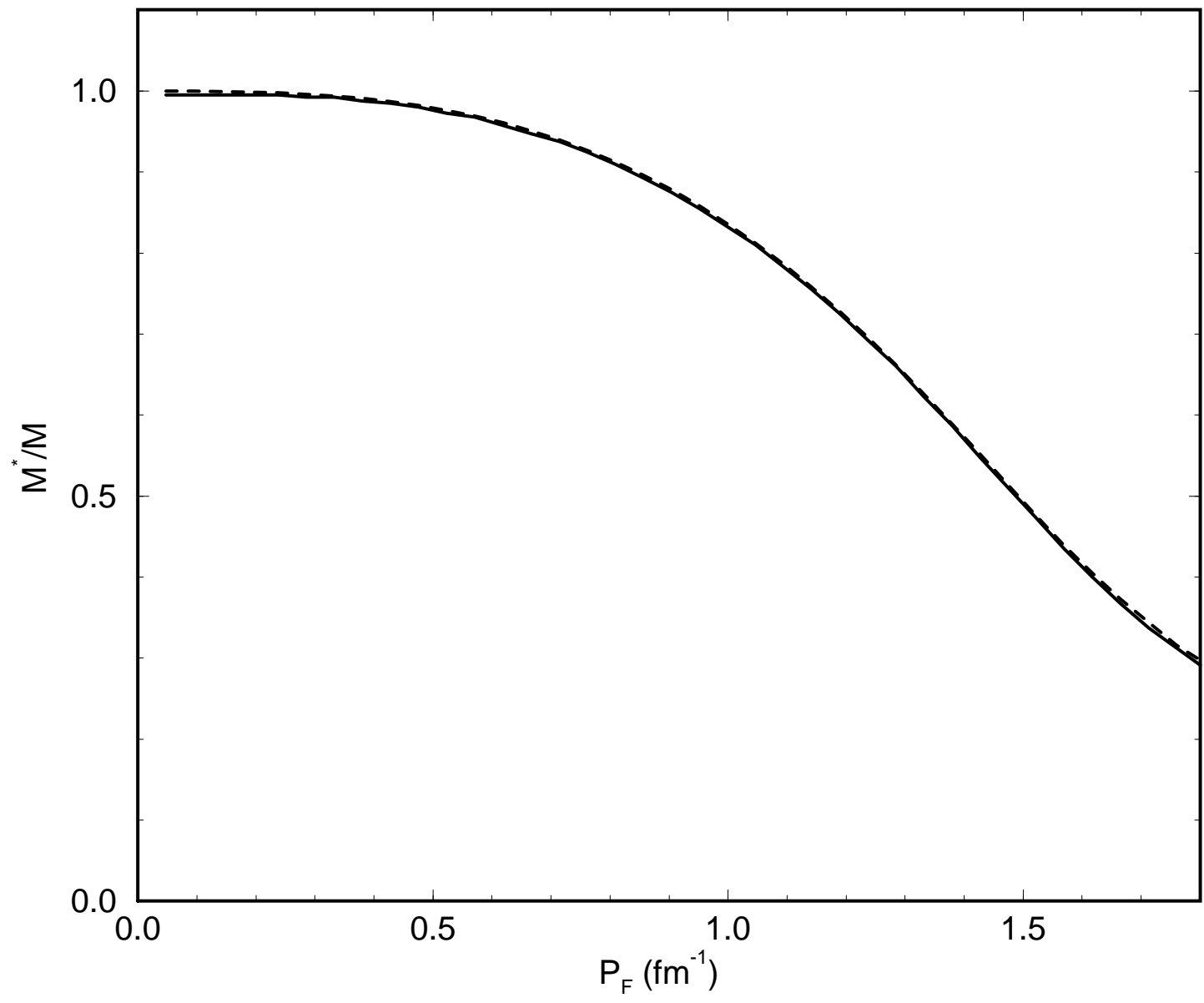
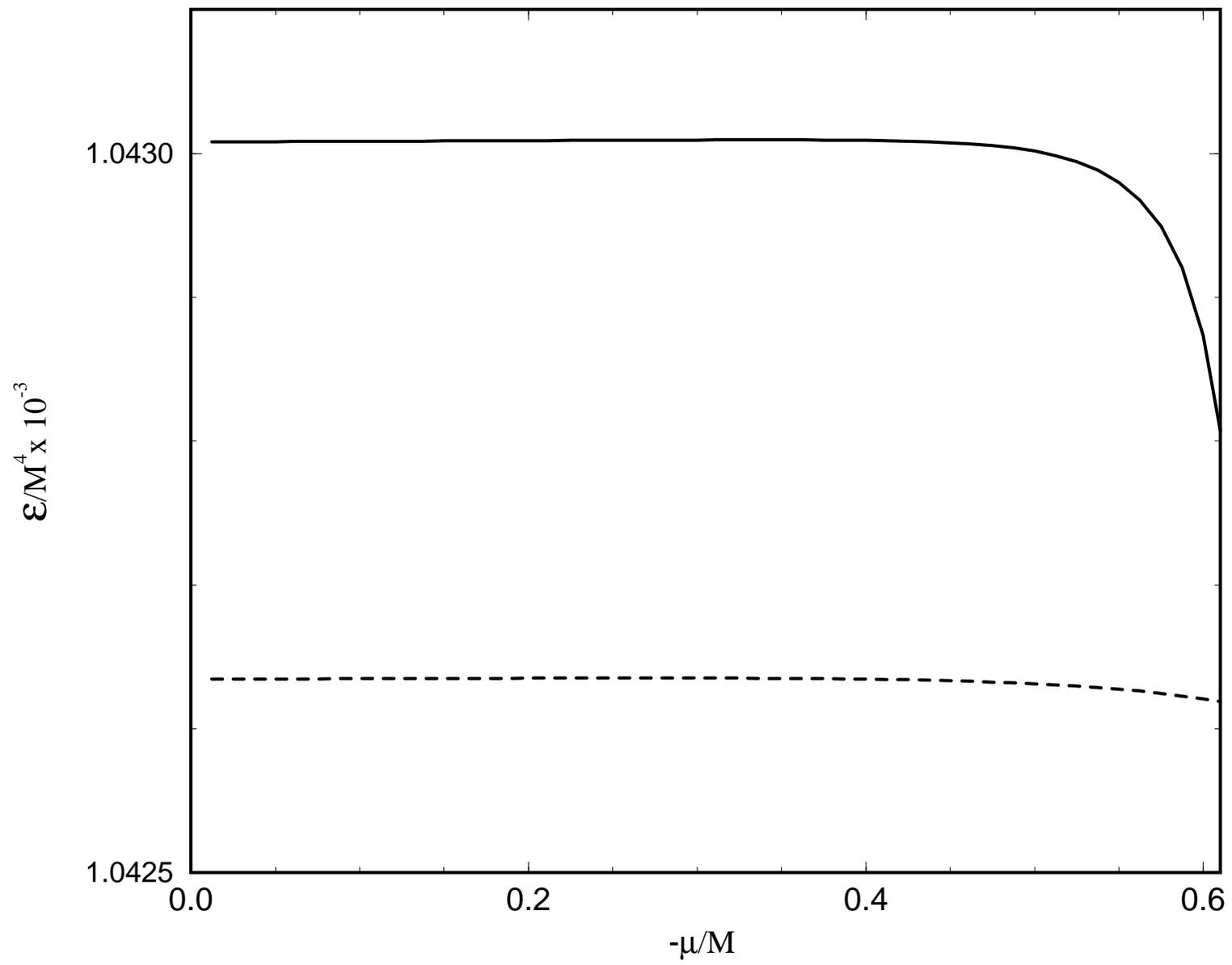
